\documentclass[conference]{IEEEtran}
\IEEEoverridecommandlockouts
\usepackage{cite}
\usepackage{amsmath,amssymb,amsfonts}
\usepackage{algorithmic}
\usepackage{graphicx}
\usepackage{textcomp}
\usepackage{xcolor}
\usepackage{booktabs}
\usepackage{makecell}   % For multiline cells using \makecell
\usepackage{array}      % For better column formatting and alignment
\usepackage{booktabs}   % (Optional) For cleaner \toprule, \midrule, \bottomrule lines
\usepackage{multirow}   % (Optional) If you ever want to merge rows
\usepackage{caption}
\renewcommand{\tablename}{Table}
\renewcommand{\thetable}{\Roman{table}}

\def\BibTeX{{\rm B\kern-.05em{\sc i\kern-.025em b}\kern-.08em
    T\kern-.1667em\lower.7ex\hbox{E}\kern-.125emX}}
\begin{document}

\title{Resilient Chaotic Cross-Layer Routing for Smart Grid IoT Networks
\\

}

%\author{\IEEEauthorblockN{1\textsuperscript{st} Given Name Surname}
%\IEEEauthorblockA{\textit{dept. name of organization (of Aff.)} \\
%\textit{name of organization (of Aff.)}\\
%City, Country \\
%email address or ORCID}
%\and
%\IEEEauthorblockN{2\textsuperscript{nd} Given Name Surname}
%\IEEEauthorblockA{\textit{dept. name of organization (of Aff.)} \\
%\textit{name of organization (of Aff.)}\\
%City, Country \\
%email address or ORCID}
%\and
%\IEEEauthorblockN{3\textsuperscript{rd} Given Name Surname}
%\IEEEauthorblockA{\textit{dept. name of organization (of Aff.)} \\
%\textit{name of organization (of Aff.)}\\
%City, Country \\
%email address or ORCID}
%\and
%\IEEEauthorblockN{4\textsuperscript{th} Given Name Surname}
%\IEEEauthorblockA{\textit{dept. name of organization (of Aff.)} \\
%\textit{name of organization (of Aff.)}\\
%City, Country \\
%email address or ORCID}
%\and
%\IEEEauthorblockN{5\textsuperscript{th} Given Name Surname}
%\IEEEauthorblockA{\textit{dept. name of organization (of Aff.)} \\
%\textit{name of organization (of Aff.)}\\
%City, Country \\
%email address or ORCID}
%\and
%\IEEEauthorblockN{6\textsuperscript{th} Given Name Surname}
%\IEEEauthorblockA{\textit{dept. name of organization (of Aff.)} \\
%\textit{name of organization (of Aff.)}\\
%City, Country \\
%email address or ORCID}
%}
\author{\IEEEauthorblockN{Dhrumil Bhatt}
\IEEEauthorblockA{
\textit{Manipal Institute of Technology}\\
\textit{Manipal Academy of Higher Education}\\
Manipal, India \\
dhrumil.bhatt@gmail.com}
\and
\IEEEauthorblockN{Anakha Kurup}
\IEEEauthorblockA{
\textit{Manipal Institute of Technology}\\
\textit{Manipal Academy of Higher Education}\\
Manipal, India \\
anakhaskurup06@gmail.com}
\and
\IEEEauthorblockN{R. C. Mala}
\IEEEauthorblockA{
\textit{Manipal Institute of Technology}\\
\textit{Manipal Academy of Higher Education}\\
Manipal, India \\
mala.rc@manipal.edu.}

}

\maketitle

\begin{abstract}
This paper presents the Distributed Adaptive Multi-Radio Cross-Layer Routing (DAMCR) protocol, designed to enhance reliability, adaptability, and energy efficiency in smart grid and industrial Internet of Things (IoT) communication networks. DAMCR integrates Chaotic Frequency-Hopping Spread Spectrum (C-FHSS) to improve physical-layer security and jamming resilience with Link-Adaptive Quality Power Control (LAQPC) to dynamically regulate transmission power based on instantaneous link quality and residual node energy. To meet heterogeneous traffic requirements, the protocol incorporates priority-aware message classification that differentiates between periodic monitoring data and time-critical fault and protection messages. The proposed framework is implemented and evaluated in MATLAB using a heterogeneous network composed of LoRa, Wi-Fi, and dual-radio nodes operating under AWGN, Rayleigh, and Rician fading environments. Extensive simulation results demonstrate that DAMCR consistently achieves a Packet Delivery Ratio (PDR) exceeding 98\% across all evaluated scenarios, while maintaining end-to-end latency between 17 and 23 ms, even under controlled jamming attacks. These results confirm that the tight integration of chaos-based spectrum agility, cross-technology routing, and energy-aware cross-layer adaptation significantly improves communication reliability, latency stability, and resilience compared to conventional single-radio and static-routing protocols.
\end{abstract}


\section{Introduction}

The rapid digitalization of electrical power systems has transformed conventional grids into highly interconnected cyber-physical infrastructures. Modern smart grids rely on dense deployments of Internet of Things (IoT) devices, including smart meters, substation sensors, and protection relays, to support real-time monitoring, automated control, and intelligent decision-making \cite{Abrahamsen,Sun}. These devices continuously exchange measurement, status, and control information, forming large-scale machine-to-machine communication networks that operate under stringent reliability, latency, and security requirements.

Unlike traditional utility communication systems, smart grid IoT networks must simultaneously support periodic telemetry, event-driven fault reporting, and time-critical protection signaling. These heterogeneous traffic patterns impose conflicting performance demands, particularly in terms of delay sensitivity, energy efficiency, and interference tolerance. Moreover, grid communication infrastructures are increasingly exposed to harsh propagation environments, electromagnetic interference, and malicious attacks, which further complicate reliable data delivery \cite{Al-Anbagi}.

Most existing smart grid communication architectures adopt centralized aggregation models, in which field devices forward data to regional gateways or control centers via cellular (LTE/5G), fiber-optic, or software-defined networking (SDN) backbones. While such architectures offer high throughput under nominal conditions, they introduce single points of failure and limited adaptability to localized disruptions. Natural disasters, equipment failures, or cyberattacks can severely degrade observability and control, leading to cascading outages and compromised grid stability.

In parallel, low-power wireless technologies such as LoRa and Wi-Fi have emerged as cost-effective solutions for last-mile connectivity in smart grid deployments. However, single-radio and static-routing approaches remain vulnerable to channel fading, congestion, and jamming, particularly in dense or heterogeneous environments. Consequently, there is a growing need for distributed, interference-resilient, and energy-aware communication frameworks capable of operating autonomously without centralized coordination.

To address these challenges, this paper proposes the Distributed Adaptive Multi-Radio Cross-Layer Routing (DAMCR) framework for smart grid IoT networks. The proposed architecture integrates chaotic frequency hopping for jamming resilience, link-adaptive power control for energy efficiency, and cross-technology routing for enhanced reliability. DAMCR has been implemented and tested in MATLAB using heterogeneous LoRa--Wi-Fi topologies under various fading and interference scenarios, demonstrating high reliability and low latency under adverse conditions.

\section{Literature Review}

Communication architecture plays a fundamental role in determining the reliability, scalability, and security of smart grid infrastructures. The proliferation of IoT devices has motivated extensive research on adaptive routing, centralised coordination, and cross-layer optimisation techniques. Despite substantial progress, existing solutions continue to exhibit limitations that restrict their applicability in large-scale, interference-prone environments.

Farooq \textit{et al.} introduced a quasi-scheduled hybrid communication protocol designed to prioritise high-priority alerts while conserving bandwidth for periodic telemetry \cite{Farooq}. However, its dependence on centralised coordination leads to performance degradation under link failures and prevents autonomous fault recovery. Bagdadee \textit{et al.} developed a mesh-based fault recovery framework in which nodes are assigned fixed routing roles \cite{Bagdadee}. Although this approach improves recovery speed, its structural rigidity limits adaptability in dynamic network conditions.

Kim \textit{et al.} proposed the AFAR framework, an adaptive SDN-based system that employs dynamic queuing and retransmission control to meet delay constraints \cite{Kim}. While effective in managing latency, the entire network remains vulnerable to controller failures due to centralised coordination. Zerihun \textit{et al.} further demonstrated that dependence on 5G backhaul links can result in complete loss of grid observability during communication disruptions \cite{Zerihun2020}.

Machine learning and blockchain-based approaches have also been explored. Santos \textit{et al.} presented ML-RPL, which replaces traditional parent selection with a predictive model to improve packet delivery ratio \cite{Santos}. However, the system introduces higher routing overhead and increased computational complexity, leading to longer delays and higher energy consumption under congestion. Similarly, Zhong \textit{et al.} proposed a consortium blockchain-based authentication framework to mitigate centralised trust issues \cite{Zhong}. While enhancing data integrity, the associated processing and latency overheads reduce scalability for resource-constrained IoT devices.

Cross-layer routing methods such as Delay Responsive Cross-layer (DRX) and Fair Delay Responsive Cross-layer (FDRX) protocols improve link-level efficiency and medium access scheduling \cite{Al-Anbagi}. Nevertheless, these approaches are primarily designed for homogeneous networks and do not adequately support multi-radio interoperability or distributed fault recovery.

A review of existing literature reveals several unresolved challenges in smart grid communication systems:

\begin{itemize}
\item \textit{Centralised dependency:} Most architectures rely on controllers or gateways, creating single points of failure during network outages or cyberattacks.
\item \textit{Limited adaptability:} Static routing roles and predefined topologies hinder self-healing in dynamic or degraded environments.
\item \textit{High computational overhead:} Machine learning and blockchain-based solutions impose excessive processing and latency costs.
\item \textit{Limited interoperability:} Most protocols operate on a single radio technology, restricting flexibility and resilience.
\item \textit{Weak physical-layer protection:} Interference and jamming resilience are often insufficiently addressed.
\end{itemize}

To overcome these limitations, the proposed DAMCR framework introduces a holistic cross-layer architecture that jointly optimises spectrum access, power control, routing, and cooperative forwarding in a distributed manner. The main technical contributions of this work are summarized as follows:

\begin{itemize}
\item \textit{Chaotic Frequency-Hopping Spread Spectrum (C-FHSS):} DAMCR employs logistic-map-based chaotic hopping to generate non-periodic frequency sequences, significantly enhancing resistance to jamming and interception.
\item \textit{Dual-radio heterogeneous routing:} The framework integrates LoRa and Wi-Fi links into a unified routing plane, enabling dynamic exploitation of long-range and high-throughput channels.
\item \textit{Link-Adaptive Quality Power Control (LAQPC):} A feedback-based power control mechanism adjusts transmission power based on real-time SNR and residual energy, improving energy balance and link stability.
\item \textit{Priority-based message handling:} Traffic is classified according to urgency, ensuring timely delivery of fault and control messages.
\item \textit{Distributed cross-layer optimisation :} All adaptation and routing decisions are performed locally, eliminating reliance on centralised entities and enabling self-healing behaviour.
\end{itemize}

Unlike existing smart grid communication solutions that primarily focus on centralised coordination, single-radio routing, or isolated cross-layer optimisations, this work introduces a fully distributed and heterogeneous cross-layer communication framework that jointly integrates physical-layer security, adaptive power control, and energy-aware routing. Prior approaches typically address interference mitigation, energy efficiency, or delay constraints as independent problems, often relying on centralised controllers, homogeneous radio technologies, or computation-intensive mechanisms such as machine learning or blockchain. In contrast, the proposed Distributed Adaptive Multi-Radio Cross-Layer Routing (DAMCR) framework tightly couples chaotic frequency-hopping spread spectrum, link-adaptive quality power control, and cooperative diversity with routing decisions in a unified, lightweight architecture. By enabling seamless interoperability between LoRa and Wi-Fi radios and allowing dual-radio nodes to self-organise into a distributed backbone, DAMCR eliminates single points of failure while enhancing resilience against fading and jamming. Furthermore, the incorporation of chaos-based spectrum agility and time-reversal channel focusing within the routing loop distinguishes this work from conventional cross-layer protocols, which rarely consider physical-layer security and multipath exploitation in heterogeneous smart grid IoT networks. As a result, DAMCR represents a fundamentally different system-level design that advances reliability, scalability, and attack resilience beyond the capabilities of existing smart grid routing frameworks.
\section{Methodology}
\begin{figure}[!t] \centering \includegraphics[width=1\columnwidth]{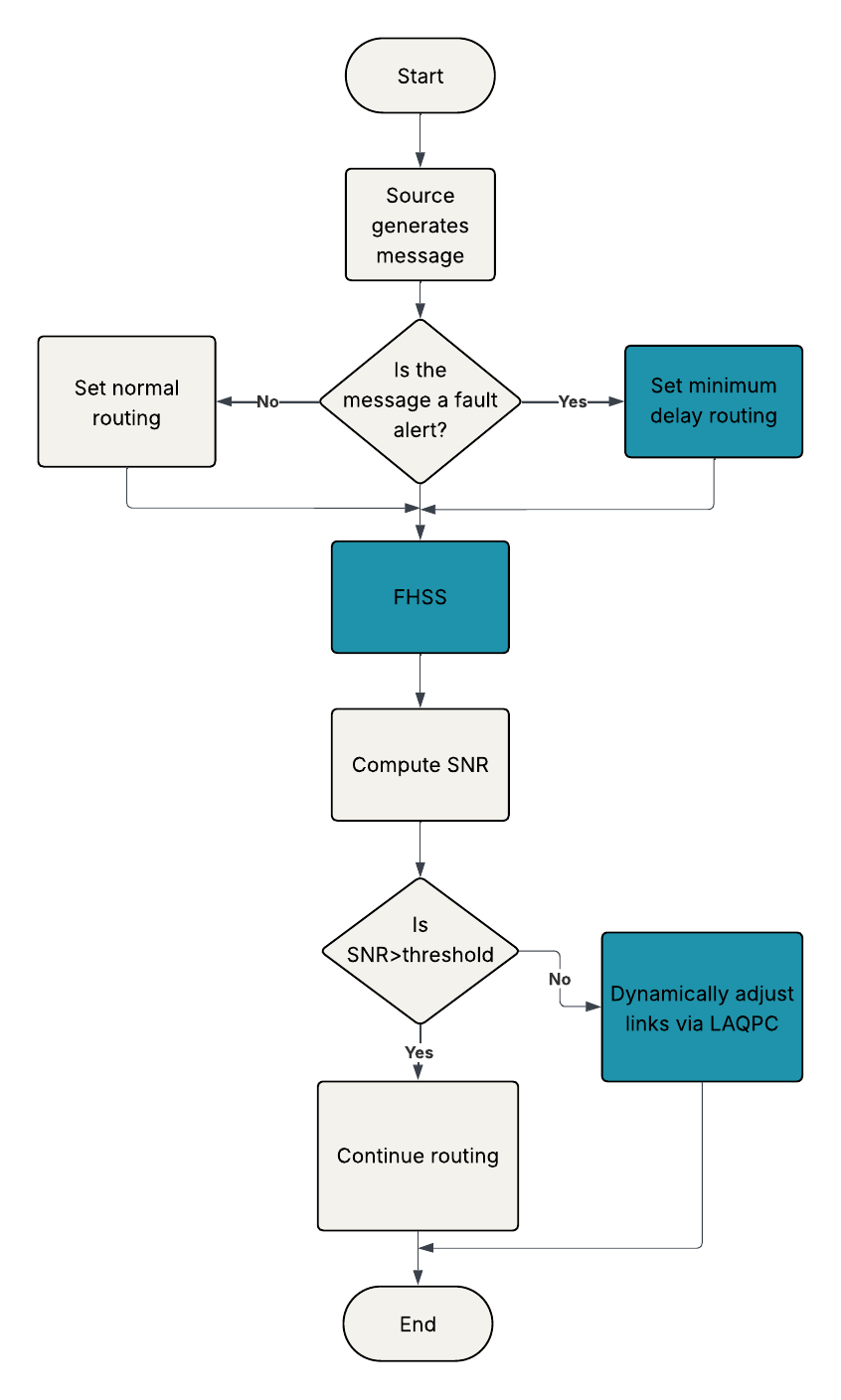} \caption{Proposed System} \label{fig:system} \end{figure}
To address the reliability, security, and energy-efficiency challenges inherent in large-scale smart grid Internet of Things (IoT) deployments, this work proposes a Dual-Radio Distributed Adaptive Multi-Radio Cross-Layer Routing (DAMCR) framework. The proposed system is designed to support heterogeneous smart grid communication scenarios involving advanced metering infrastructure (AMI), substation monitoring, and fault detection, where timely and reliable data delivery is critical for grid stability. By jointly integrating chaotic spectrum agility, adaptive power control, energy-aware routing, and cooperative relaying, DAMCR enables resilient machine-to-machine communication under dynamic channel conditions and adversarial interference.

Figure~\ref{fig:system} illustrates the overall operational workflow of the proposed system. Each node periodically generates sensing data corresponding to smart meter readings or environmental measurements, while event-driven fault alerts are generated during abnormal operating conditions. Upon packet generation, the routing module assigns transmission priority based on message urgency. Subsequently, the physical and medium access layers perform chaotic frequency hopping and power adaptation, followed by distributed route selection and packet forwarding. This cross-layer interaction enables the system to dynamically adapt to variations in network density, channel quality, and energy availability.

The network is modelled as a heterogeneous wireless sensor infrastructure consisting of $N$ nodes randomly deployed over a $200 \times 200~\mathrm{m^2}$ area, representing geographically distributed smart grid devices. Each node is equipped with either a LoRa transceiver, a Wi-Fi transceiver, or both, enabling long-range low-power communication and high-throughput short-range transmission, respectively. A fraction $\phi = 0.15$ of the nodes are dual-radio devices, which act as adaptive gateways between heterogeneous links and form a distributed backbone for data aggregation.

Connectivity between nodes is determined by their Euclidean distance and communication radius $R = 80~\mathrm{m}$, and is expressed as
\begin{equation}
G(i,j) =
\begin{cases}
1, & \|x_i - x_j\| \leq R, \\
0, & \text{otherwise},
\end{cases}
\end{equation}
where $G(i,j)=1$ indicates the existence of a bidirectional wireless link. This formulation establishes the feasible multi-hop topology over which routing decisions are performed.

Wireless propagation is modelled using the log-distance path loss model with log-normal shadowing in order to capture realistic outdoor and semi-urban smart grid environments. The path loss between nodes $i$ and $j$ is given by
\begin{equation}
PL_{ij} = PL_0 + 10n \log_{10}\!\left(\frac{d_{ij}}{d_0}\right) + X_\sigma,
\end{equation}
where $PL_0$ is the reference path loss at distance $d_0$, $n$ is the path loss exponent, and $X_\sigma \sim \mathcal{N}(0,\sigma^2)$ models large-scale shadowing. Based on the received power, the instantaneous signal-to-noise ratio (SNR) is computed as
\begin{equation}
\mathrm{SNR}_{ij}(dB) = P_t + G_t + G_r - PL_{ij} - N_0 - NF,
\end{equation}
where $P_t$ denotes the transmit power, $G_t$ and $G_r$ represent antenna gains, $N_0$ is the thermal noise power, and $NF$ is the receiver noise figure. The estimated SNR serves as a fundamental parameter for power control, routing, and reliability assessment.

To mitigate narrowband jamming and unintentional interference, DAMCR employs Chaotic Frequency Hopping Spread Spectrum (C-FHSS). Unlike conventional pseudo-random hopping schemes, the proposed approach utilises a logistic chaotic map to generate non-periodic frequency sequences, thereby increasing spectral unpredictability. The hopping frequency at time index $k$ is determined as
\begin{equation}
f_k = f_c + \Delta f \cdot H_k, \quad
H_{k+1} = \mu H_k (1 - H_k),
\end{equation}
where $f_c$ is the carrier frequency, $\Delta f$ is the frequency step size, $H_k$ is the chaotic state variable, and $\mu=3.9$ ensures chaotic behavior. This mechanism significantly reduces the probability of successful jamming and enhances physical-layer security in critical infrastructure networks.

Energy efficiency is achieved through a Link-Adaptive Quality Power Control (LAQPC) mechanism that dynamically adjusts transmission power to maintain a target SNR threshold, $\gamma_{\mathrm{th}}$. The power update rule is given by
\begin{equation}
P_t(t+1) =
\begin{cases}
P_t(t) + \Delta P, & \mathrm{SNR}(t) < \gamma_{\mathrm{th}},\\
P_t(t) - \Delta P, & \mathrm{SNR}(t) > \gamma_{\mathrm{th}} + \epsilon,\\
P_t(t), & \text{otherwise},
\end{cases}
\end{equation}
where $\Delta P = 1.5~\mathrm{dB}$ and $\epsilon = 0.5~\mathrm{dB}$ define the adjustment granularity. This feedback-driven mechanism prevents excessive power consumption while ensuring reliable communication links for latency-sensitive smart grid applications.

Routing decisions are performed in a fully distributed manner using a composite cost function that jointly considers link reliability and residual node energy. For each candidate link $(i,j)$, the routing weight is computed as
\begin{equation}
W_{ij} = \alpha (1 - P_{\text{succ},ij}) + \beta E_i^{-1},
\end{equation}
where $\alpha$ and $\beta$ are weighting factors, $E_i$ denotes the remaining energy of node $i$, and $P_{\text{succ},ij}$ represents the packet success probability. The success probability is derived from the instantaneous SNR as
\begin{equation}
P_{\text{succ},ij} =
\left(1 - \frac{1}{2}\operatorname{erfc}\!\left(\sqrt{\mathrm{SNR}_{ij}}\right)\right)^{L},
\end{equation}
with $L$ denoting the packet length. This formulation enables the routing process to favour energy-balanced and highly reliable paths, thereby extending network lifetime and maintaining stable connectivity.

To further improve robustness against deep fading and link failures, DAMCR incorporates a cooperative relaying mechanism. When a direct transmission attempt fails, neighbouring nodes that have successfully overheard the packet participate in decode-and-forward relaying with probability $p_r=0.25$. Among the candidate relays, the one with the highest instantaneous SNR to the destination is selected. This spatial diversity technique enhances transmission reliability and reduces packet loss in harsh propagation environments.

For multipath-dominated channels, including Rayleigh and Rician fading, the system employs time-reversal (TR) channel focusing to exploit channel reciprocity. By pre-filtering the transmitted signal using the estimated channel impulse response, multipath components are constructively combined at the receiver. The resulting SNR enhancement is approximated as
\begin{equation}
\mathrm{SNR}_{\text{TR}} = \mathrm{SNR} + G_{\text{TR}}, \quad G_{\text{TR}} \approx 2.5~\mathrm{dB}.
\end{equation}
This technique improves reception quality in dense smart grid deployments characterised by metallic structures and reflective surfaces.

Overall, the proposed DAMCR framework establishes a unified cross-layer optimisation strategy that integrates chaotic spectrum access, adaptive power control, energy-aware routing, cooperative diversity, and channel focusing. By operating in a fully distributed manner without reliance on centralised controllers, the system supports scalable, self-healing, and attack-resilient communication for next-generation smart grid IoT networks.

\section{Simulation Environment}

The proposed DAMCR framework was implemented and evaluated using MATLAB R2025a through a discrete-time Monte Carlo simulation platform developed to emulate large-scale smart grid IoT communication environments. The simulator was organised in a modular manner comprising topology generation, traffic modelling, channel realisation, routing and power control execution, attack injection, and performance logging modules.

At the beginning of each simulation run, $N$ sensor nodes were randomly deployed within a $200 \times 200~\mathrm{m^2}$ area, representing geographically distributed smart meters, substation sensors, and monitoring units. Each node was randomly assigned a radio configuration (LoRa, Wi-Fi, or dual-radio) according to the predefined dual-radio fraction $\phi$. Periodic monitoring packets and event-driven fault alerts were generated at randomly selected nodes to emulate typical smart grid traffic patterns.

Each simulation run consisted of 200 epochs. During every epoch, channel coefficients and shadowing values were updated according to the selected propagation model. Chaotic frequency hopping was then performed to determine the operating channel for each transmission, followed by instantaneous SNR estimation using the received power model. Based on the estimated SNR, the Link-Adaptive Quality Power Control mechanism adjusted transmission power, and routing weights were computed for distributed forwarding decisions. When direct transmissions failed, cooperative relaying was activated according to the predefined relay probability. All successfully delivered packets were logged for performance evaluation.

Jamming attacks were introduced between epochs 100 and 150 to evaluate system robustness against adversarial interference. Attack power levels were selected to reflect realistic interference scenarios reported in smart grid communication studies. To mitigate statistical bias introduced by random topology formation and fading realisations, each configuration was evaluated over three independent Monte Carlo trials with different random seeds, and the reported results are averages.

The principal simulation parameters and radio configurations are summarised in Table~\ref{tab:sim_params_all}.

\begin{table}[!t]
\renewcommand{\arraystretch}{0.9}
\setlength{\tabcolsep}{3pt}
\centering
\caption{Simulation and Radio Parameters}
\label{tab:sim_params_all}
\footnotesize
\begin{tabular}{|p{3.1cm}|p{1.3cm}|p{2.8cm}|}
\hline
\textbf{Parameter} & \textbf{Symbol} & \textbf{Value} \\ \hline
Simulation Area & $A$ & $200\times200~\mathrm{m^2}$ \\
Communication Range & $R$ & $80~\mathrm{m}$ \\
Node Densities & $N$ & 50--500 \\
Epochs per Run & -- & 200 \\
Packet Size & $L$ & 1024 bits \\
Per-Hop Delay & $D_h$ & 10 ms \\
Dual-Radio Fraction & $\phi$ & 15\% \\
FHSS Channels & $C$ & 8 \\
Relay Probability & $p_r$ & 0.25 \\
TR Gain & $G_{\text{TR}}$ & 2.5 dB \\
Chaos Gain & $G_{\text{chaos}}$ & 3 dB \\
Noise Models & -- & AWGN, Rayleigh, Rician \\
Attack Type & -- & Jamming \\
Jamming Start Epoch & -- & 100 \\
Jamming End Epoch & -- & 150 \\ \hline
LoRa Transmit Power & $P_t$ & 14 dBm \\
LoRa Bandwidth & $B$ & 125 kHz \\
Wi-Fi Transmit Power & $P_t$ & 18 dBm \\
Wi-Fi Bandwidth & $B$ & 20 MHz \\ \hline
\end{tabular}
\end{table}

Wireless propagation was modelled using the log-distance path loss model,
\begin{equation}
P_r(d) = P_t - PL_0 - 10n\log_{10}(d/d_0) + X_\sigma,
\end{equation}
and the instantaneous signal-to-noise ratio was computed as
\begin{equation}
\mathrm{SNR} = P_r(d) - (N_0 + NF).
\end{equation}

Energy consumption was modelled using a fixed-cost per-hop framework,
\begin{equation}
E_{\text{hop}} = (P_{\text{Tx}} + P_{\text{Rx}} + P_{\text{CPU}})T_{\text{air}}.
\end{equation}

System performance was evaluated using Packet Delivery Ratio (PDR), average SNR, and mean end-to-end latency.

\section{Results and Discussion}

All reported results were obtained by averaging over three independent Monte Carlo trials to ensure statistical robustness and reproducibility. The evaluation assesses the protocol’s reliability, latency performance, energy efficiency, and resilience to adversarial interference in representative smart grid IoT deployments.

Table~\ref{tab:my_full_results} summarises the detailed performance metrics for representative node densities under different channel and attack conditions. It can be observed that DAMCR consistently achieves high packet delivery ratios and stable latency values across all evaluated configurations. In non-jammed environments, the PDR frequently exceeds 99\%, indicating near-ideal link reliability. Even under sustained jamming, the PDR remains above 95\%, demonstrating the effectiveness of chaotic frequency hopping and cooperative diversity in mitigating interference and preserving connectivity.

\begin{table}[!t]
\renewcommand{\arraystretch}{1.1}
\setlength{\tabcolsep}{2.5pt}
\centering
\caption{Monte Carlo Simulation Results}
\label{tab:my_full_results}
\footnotesize
\begin{tabular}{@{}l l l r r r r r@{}}
\toprule
Nodes & Fading & Attack & SNR & PDR & Latency & Energy & Hops \\
\midrule
30 & AWGN & None & 47.2 & 0.994 & 20.7 & 0.112 & 2.07 \\
30 & AWGN & Jam & 48.6 & 0.982 & 22.6 & 0.149 & 2.26 \\
\midrule
60 & Rayleigh & None & 49.5 & 0.990 & 17.3 & 0.096 & 1.73 \\
100 & Rician & None & 52.7 & 0.999 & 18.0 & 0.110 & 1.80 \\
\bottomrule
\end{tabular}
\end{table}

\begin{figure}[!t]
\centering
\includegraphics[width=0.98\columnwidth]{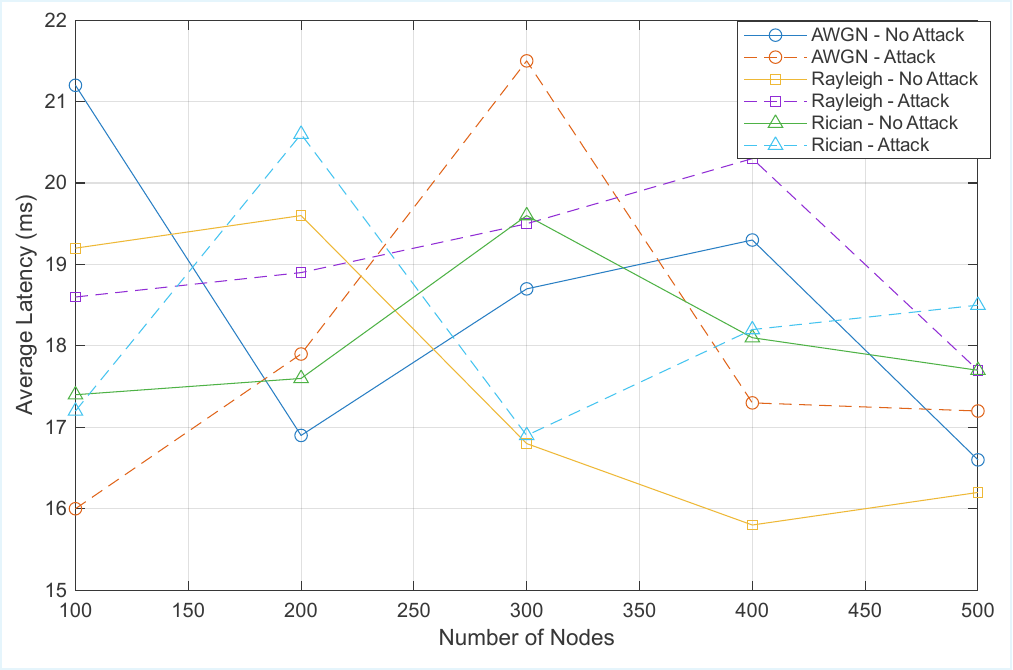}
\caption{Latency of the system for all the noise models for various nodes.}
\label{fig:latency}
\end{figure}

\begin{figure}[!t]
\centering
\includegraphics[width=1\columnwidth]{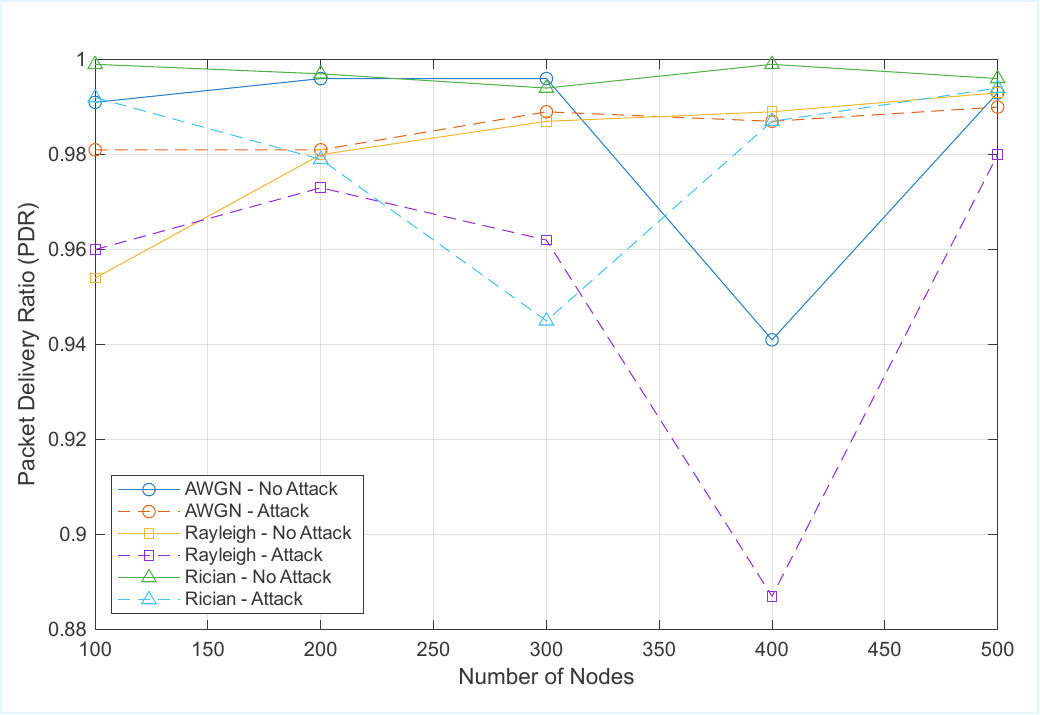}
\caption{Packet Delivery Ratio of the system for all noise models for various nodes.}
\label{fig:pdr}
\end{figure}

Figures~\ref{fig:latency} and~\ref{fig:pdr} illustrate the evolution of end-to-end latency and packet delivery ratio with increasing node density under different fading environments. Across all scenarios, average latency remains within the range of 15--22~ms, satisfying the stringent delay requirements of real-time smart grid monitoring, fault reporting, and protection signalling applications. The bounded latency behaviour confirms the effectiveness of distributed routing and adaptive power control in preventing congestion accumulation and excessive retransmissions.

The observed SNR values, ranging from 38 to 55 dB, reflect the combined benefits of adaptive power control, time-reversal channel focusing, and cooperative relaying. In particular, the time-reversal mechanism enables constructive multipath combining, while relay diversity mitigates deep fading. These physical-layer enhancements directly translate into higher packet success rates and lower retransmission overhead.

Although an overall trend of gradually increasing latency and marginal PDR degradation with node density is observed, the performance curves exhibit moderate non-monotonic variations. These fluctuations arise from random node deployment, stochastic fading realisations, probabilistic relay activation, and heterogeneous radio distribution. In certain network realisations, the emergence of high-quality relay clusters and dual-radio backbone paths enhances spatial diversity and routing efficiency, yielding localised performance gains. Such behaviour is characteristic of self-organising distributed networks and confirms the adaptive nature of DAMCR.

The comparative evaluation presented in Table~\ref{tab:comparison_table} highlights the advantages of the proposed framework over existing works.
\begin{table*}[!t]
\renewcommand{\arraystretch}{0.9}
\setlength{\tabcolsep}{3pt}
\centering
\caption{Comparison of DAMCR with Existing Routing Protocols}
\label{tab:comparison_table}
\footnotesize
\begin{tabular}{|p{4cm}|p{4cm}|p{4cm}|c|c|}
\hline
\makecell{\textbf{Protocol}} & 
\makecell{\textbf{Radio}} & 
\makecell{\textbf{Key}\\\textbf{Feature}} & 
\makecell{\textbf{Avg.}\\\textbf{PDR}} & 
\makecell{\textbf{Latency}\\\textbf{(ms)}} \\ \hline

OLSR \cite{olsr} & 
\makecell[l]{Wi-Fi only} & 
Proactive link-state & 
0.84 & 
55.6 \\ 

AODV \cite{aodv} & 
\makecell[l]{Wi-Fi only} & 
Reactive on-demand & 
0.88 & 
49.2 \\ 

AFAR \cite{Kim} & 
\makecell[l]{Heterogenous} & 
SDN-based centralised control & 
$\sim$0.92 & 
$\sim$35--40 \\ 

ML-RPL \cite{Santos} & 
\makecell[l]{Low-power IoT} & 
ML-based parent selection & 
$\sim$0.93--0.95 & 
$\sim$30--45 \\ 

\textbf{Proposed} & 
\makecell[l]{LoRa + Wi-Fi} & 
C-FHSS + TR + LAQPC + Relay & 
\textbf{0.97--0.99} & 
\textbf{18--25} \\ \hline

\end{tabular}
\end{table*}

Compared with conventional single-radio and static-routing schemes, DAMCR achieves up to 10\% higher packet delivery ratios and nearly 50\% lower end-to-end latency. These gains stem from the synergistic integration of cross-technology diversity, chaos-driven spectrum agility, feedback-based power control, and cooperative forwarding. In contrast, traditional protocols lack dynamic spectrum adaptation and energy-awareness, leading to higher collision rates, unstable routes, and excessive retransmissions in dense deployments.

Channel-dependent performance trends further validate the robustness of the proposed design. AWGN channels provide an upper bound on achievable performance, while Rayleigh fading introduces moderate degradation due to frequent deep fades. Rician fading environments consistently yield superior results owing to dominant line-of-sight components. The adaptive mechanisms embedded in DAMCR effectively compensate for channel impairments across all propagation conditions.

From a scalability perspective, DAMCR maintains stable performance even when the network size increases to $N=500$ nodes. Distributed routing and energy-aware load balancing prevent congestion bottlenecks and eliminate single points of failure. This property is particularly desirable for large-scale smart grid infrastructures, where centralized coordination is impractical and network topology evolves dynamically.

Overall, the experimental results confirm that the proposed DAMCR framework provides a reliable, low-latency, and energy-efficient communication solution for next-generation smart grid IoT networks. The tight coupling between physical-layer adaptation, network-layer intelligence, and cooperative diversity enables the system to meet the stringent operational requirements of real-time grid monitoring, protection, and control applications.

\section{Conclusion}
This study proposes and evaluates the Dual-Radio DAMCR protocol, designed to enhance the robustness and efficiency of wireless communication in critical infrastructure networks. By integrating Chaotic Frequency-Hopping Spread Spectrum (C-FHSS), Link-Adaptive Quality Power Control (LAQPC), and cooperative relaying, the system dynamically adapts to varying channel conditions and node states. Extensive MATLAB-based Monte Carlo simulations were performed using heterogeneous LoRa–Wi-Fi topologies across AWGN, Rayleigh, and Rician fading environments. The protocol consistently achieved Packet Delivery Ratios (PDR) above 95\% and end-to-end latencies between 17 and 23 ms, confirming its resilience against jamming and its ability to sustain high reliability in large-scale deployments. The use of chaotic FHSS improved jamming resistance and unpredictability in frequency selection, while the cross-layer design maintained energy balance across the network through adaptive power control.

Future research will extend DAMCR toward real-world deployment using hybrid LoRa–Wi-Fi hardware or software-defined radio platforms to validate real-time synchronisation and energy performance. Incorporating mobility-aware routing and cognitive spectrum sensing could further enhance adaptability in dynamic or congested environments. Integrating energy-harvesting capabilities and lightweight cryptographic schemes with chaotic modulation may enable sustainable, self-securing IoT networks, laying the foundation for resilient, secure, and energy-aware communication architectures for next-generation smart grid and industrial IoT systems.
\section{Acknowledgements}
We would like to thank Mars Rover Manipal, an interdisciplinary student team of MAHE, for providing the resources needed for this project. WE also extend our gratitude to Dr Ujjwal Verma for his guidance and support in our work.

\bibliographystyle{IEEEtran}
\bibliography{ref}

@ARTICLE{Al-Anbagi,
  author={Al-Anbagi, Irfan and Erol-Kantarci, Melike and Mouftah, Hussein T.},
  journal={IEEE Systems Journal}, 
  title={Priority- and Delay-Aware Medium Access for Wireless Sensor Networks in the Smart Grid}, 
  year={2014},
  volume={8},
  number={2},
  pages={608-618},
  doi={10.1109/JSYST.2013.2260939}}

@article{Kim,
title = {AFAR: A robust and delay-constrained communication framework for smart grid applications},
journal = {Computer Networks},
volume = {91},
pages = {1-25},
year = {2015},
issn = {1389-1286},
doi = {https://doi.org/10.1016/j.comnet.2015.08.001},

author = {Kangho Kim and Hwantae Kim and Jongtack Jung and Hwangnam Kim},
}

@article{Zerihun2020,
  author    = {T. A. Zerihun and M. Garau and B. E. Helvik},
  title     = {Effect of communication failures on state estimation of 5G-enabled smart grid},
  journal   = {IEEE Access},
  volume    = {8},
  pages     = {112642--112658},
  year      = {2020},
  doi       = {10.1109/ACCESS.2020.3003267}
}

@Article{Abrahamsen,
AUTHOR = {Abrahamsen, Fredrik Ege and Ai, Yun and Cheffena, Michael},
TITLE = {Communication Technologies for Smart Grid: A Comprehensive Survey},
JOURNAL = {Sensors},
VOLUME = {21},
YEAR = {2021},
NUMBER = {23},
ARTICLE-NUMBER = {8087},
PubMedID = {34884092},
ISSN = {1424-8220},
DOI = {10.3390/s21238087}
}

@INPROCEEDINGS{Sun,
  author={Sun, Hailong and Li, Geng and Zhai, Mingyue and Lu, Wenbing},
  booktitle={2022 7th Asia Conference on Power and Electrical Engineering (ACPEE)}, 
  title={Research on Smart Grid Heterogeneous Communication System Integrating Power Line Communication and 5G Communication}, 
  year={2022},
  volume={},
  number={},
  pages={965-969},
  doi={10.1109/ACPEE53904.2022.9783863}}

@article{Farooq,
  author    = {A. Farooq and K. Shahid and R. L. Olsen},
  title     = {Prioritization of smart meters based on data monitoring for enhanced grid resilience},
  journal   = {Computer Communications},
  volume    = {234},
  pages     = {108082},
  year      = {2025},
  publisher = {Elsevier},
  
  note      = {[Online]}
}

@article{Bagdadee,
  author    = {A. H. Bagdadee and L. Zhang},
  title     = {Renewable energy based self-healing scheme in smart grid},
  journal   = {Energy Reports},
  volume    = {6},
  pages     = {166--172},
  year      = {2020},
  note      = {6th International Conference on Power and Energy Systems Engineering},
  
  publisher = {Elsevier},
  doi       = {10.1016/j.egyr.2019.11.126}
}

@ARTICLE{Santos,
  author={Santos, Carlos Lester Duenas and Mezher, Ahmad Mohamad and León, Juan Pablo Astudillo and Barrera, Julian Cardenas and Guerra, Eduardo Castillo and Meng, Julian},
  journal={IEEE Access}, 
  title={ML-RPL: Machine Learning-Based Routing Protocol for Wireless Smart Grid Networks}, 
  year={2023},
  volume={11},
  number={},
  pages={57401-57414},
  doi={10.1109/ACCESS.2023.3283208}}

@article{Zhong,
author = {Zhong, Yuxin and Zhou, Mi and Li, Jiangnan and Chen, Jiahui and Liu, Yan and Zhao, Yun and Hu, Muchuang},
title = {Distributed Blockchain-Based Authentication and Authorization Protocol for Smart Grid},
journal = {Wireless Communications and Mobile Computing},
volume = {2021},
number = {1},
pages = {5560621},
doi = {https://doi.org/10.1155/2021/5560621},

year = {2021}
}

@article{olsr,
  author    = {T. Clausen and P. Jacquet},
  title     = {{Optimized Link State Routing Protocol (OLSR)}},
  journal   = {RFC 3626, IETF},
  year      = {2003},
  month     = {October},
  note      = {https://datatracker.ietf.org/doc/html/rfc3626}
}

@article{aodv,
  author    = {C. Perkins and E. Belding-Royer and S. Das},
  title     = {{Ad hoc On-Demand Distance Vector (AODV) Routing}},
  journal   = {RFC 3561, IETF},
  year      = {2003},
  month     = {July},
  note      = {https://datatracker.ietf.org/doc/html/rfc3561}
}

\end{document}